\documentstyle[aps,prl,multicol,epsf]{revtex}
\def\beq{\begin{equation}}
\def\eeq{\end{equation}}
\begin{document}                
\title{Glass Formation in a Periodic Long-Range Josephson Array}
\author{P. Chandra$^1$, M.V. Feigelman$^2$ and L.B. Ioffe$^{2,3}$}
\address{$^1$NEC Research Institute, 4 Independence Way, Princeton NJ
08540}
\address{$^2$Landau Institute for Theoretical Physics, Moscow, RUSSIA}
\address{$^3$Department of Physics, Rutgers University, Piscataway, NJ 08855}
\maketitle
\begin{abstract}
We present an analytic study of a  dynamical instability in a periodic
long-range Josephson
array frustrated by a weak transverse field.
This glass transition is characterized by a diverging
relaxation time and a jump in the Edwards-Anderson order parameter;
it is {\sl not} accompanied by a coinciding static transition.

\end{abstract}

\begin{multicols}{2}

Glass formation in the {\sl absence} of intrinsic disorder is a long-standing
problem.
Although vitrification is ubiquitous,
a minimalist microscopic model of this phenomenon remains the subject of active
discussion.\cite{activity}
Because glass formation is a dynamical transition that is not necessarily
accompanied by a static one, it lies outside the framework of the Landau
theory.
Furthermore this glass transition leads to a low-temperature phase with broken
ergodicity {\sl without} the selection of a {\sl unique} state.
A successful phenomenological theory should describe the ``partitioning'' of
phase space below the transition temperature $(T_G)$ into an exponential
number of metastable states; specifically it should explain how the system
becomes ``stuck'' at $T_G$ in one of these states that is separated from the
others by barriers that scale with the system size.

Unfortunately a basic theory of
glass formation has not yet been found.
Several microscopic non-random models have been proposed; most were studied via
a mapping to disordered systems \cite{marinari}.
Recently possible glassiness in the absence of disorder has been studied
in a periodic long-range Josephson array using a direct, analytic
approach
\cite{us};
furthermore this system may be realized experimentally.
An analysis of its static behavior indicates a first-order transition into a
low-temperature phase characterized by an extensive number of states
separated by infinite barriers.
In this Letter we continue the study of this system and show that it displays
a true dynamical instability that {\sl precedes} the static transition, as
expected in a glass.\cite{activity}

The proposed  array is a stack of two mutually perpendicular sets of $N$
parallel wires with Josephson junctions at each node (Figure 1)  that is
placed in an external tranverse field.
The classical thermodynamic variables of this system are the superconducting
phases associated with each wire.
Here we shall assume that the Josephson couplings are sufficiently small so
that the induced fields are negligible in comparison with $H$.
We can therefore describe the array by the Hamiltonian
\beq
{\cal H} = - \sum_{m,n}^{2N} s_m^{*} {\cal J}_{mn} s_n
\label{H}
\eeq
where ${\cal J}_{mn}$ is the coupling matrix
\beq
\hat{\cal  J} = \left( \begin{array}{cc}
0 & \hat{J} \\
\hat{J}^\dagger & 0
\end{array}
\right)
\label{J}
\eeq
with $J_{jk} = \frac{J_0}{\sqrt{N}} \exp(2\pi i \alpha jk /N)$ and $1 \! \leq
\! (j,k) \! \leq \! N$ where $j(k)$ is the index of the horizontal (vertical)
wires;
$s_m = e^{i\phi_m}$  where the $\phi_m$ are the superconducting phases of the
$2N$ wires.
Here we have introduced the flux per unit strip, $\alpha = NHl^2/\phi_0$,
where   $l$ is the inter-node spacing and $\phi_0$ is the flux quantum;
the normalization has been chosen so that $T_G$ does not scale with
$N$.

\begin{figure}
\vspace{-.35in}
\centerline{\epsfxsize=6cm \epsfbox{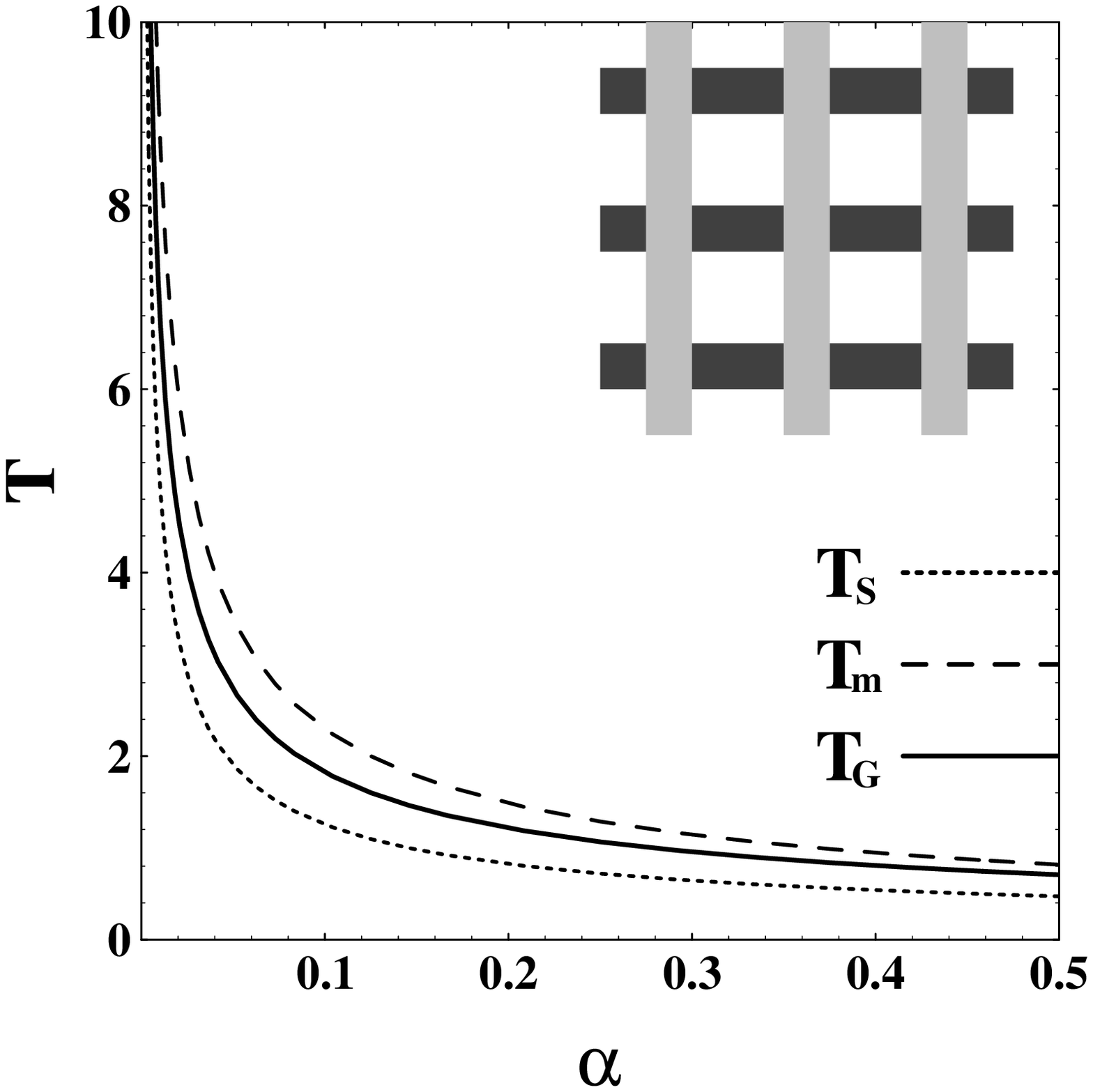}}
\vspace{-.35in}
Fig 1.
{The phase diagram of the array (inset) where $T_G$ indicates the temperature
associated with the dynamical instability discussed in this Letter, $T_S$ is
the speculated equilibrium transition temperature and $T_m$ is the
``superheating'' temperature where the low-temperature metastable states cease
to exist.}
\end{figure}
\vspace{-.1in}

Because every horizontal (vertical) wire is linked to every vertical
(horizontal) wire, the number of nearest neighbors ($z$) in this model is $N$;
we can therefore study it with a mean-field approach.
For $\frac{1}{N} \ll \alpha < 1$ the number of low-temperature metastable
solutions is extensive \cite{us}.
Furthermore this degeneracy develops simultaneously with the instability
of the paramagnetic phase; at this temperature interactions do not favor a
particular state
and, because $z \sim N$,
the barriers separating these low-temperature metastable solutions
are effectively infinite.
The static response displays {\sl no} soft mode instability but indicates a
jump in the Edwards-Anderson order parameter in the vicinity of $T_0 =
\frac{J_0}{\sqrt{\alpha}}$.

Before presenting our quantitative treatment of the dynamical behavior
of this array, we discuss the qualitative picture of
the glass transition that emerges from our results (cf. Fig. 1).
As $T$ approaches $T_m^+$, where $T_m^+ \sim T_0$, there
appear a number of metastable states in addition to the paramagnetic
free-energy minimum; most likely they are energetically unfavorable and thus
do not ``trap'' the system upon cooling from high temperatures.
As $T \rightarrow T_G^+$, the paramagnetic minimum is ``subdivided'' into an
extensive number of degenerate metastable states separated by effectively
infinite barriers, and the system is dynamically localized into one of them.
Qualitatively, in the interval $T_m > T > T_G$ there appear many
local minima in the vicinity of the paramagnetic state separated by {\sl
finite} barriers; these barriers increase continuously and become
infinite at $T = T_G$.
Each of these minima is characterized by a finite ``site magnetization'' $m_i
= \langle s_i \rangle_T$  where``site'' refers to a wire.
When $T > T_G$ thermal fluctuations average over many states so that $\langle
m_i \rangle \equiv 0$.
At $T=T_G$ the system is localized in one metastable state and there is an
associated jump in the Edwards-Anderson order parameter, $\left (q =
\frac{1}{N}\sum_i \langle m_i\rangle^2 \right)$.
The low-temperature phase is characterized by a finite $q$ and by the presence
of a memory, $\lim_{t' \rightarrow \infty} \Delta(t,t') \neq 0$ where $\Delta
(t,t')$ is the anomalous response.
We expect that at $T = T_G$, the metastable states are degenerate and thus
there can be no thermodynamic selection.
However at lower temperatures interactions will probably break this degeneracy
and select a subset of this manifold;  then we expect an
($t \rightarrow \infty$) equilibrium first-order transition ($T_S$) which
should be accompanied by a jump in the local magnetization.
In order to observe this transition at $T_S$  the array must be
equilibrated
on a time-scale ($t_W$) longer than that ($t_A$) necessary to overcome
the barriers separating its metastable states; $t_A$ scales
exponentially with the number of wires in the array.
Thus the equilibrium transition at $T_S$ is observable {\sl only}
if $t_W \rightarrow \infty$ {\sl before} the thermodynamic limit
($N \rightarrow \infty$) is taken; in the opposite order of limits
only the dynamical transition occurs.

We now begin a more quantitative analysis of the dynamic instability
in this periodic array.
Because our focus is on the long-time behavior of this system,
we expect the details of the single-spin dynamics to be irrelevant;
we therefore choose to study the simplest form, namely that of
soft spins with Langevin relaxational dynamics. More specifically,
we introduce a ``potential'', $V(S_i) = V_0 (|S|^2 - 1)^2$, at each
wire which constrains the magnitude of each spin, $|S_i| \approx 1$,
and assume the equations of motion
\begin{eqnarray}
&\tau_b \dot{S_i}& = -\frac{1}{T} \frac{\partial ({\cal H} + V)}{\partial
S_i} + \zeta_i
\label{langevin}\\
\langle &\zeta_i(t)&  \zeta_j (t')\rangle = 2\tau_b \delta(t - t')
\end{eqnarray}
where $\tau_b$ is a microscopic time-scale.
The dynamics (\ref{langevin}) reproduces the dynamics of the overdamped
Josephson junctions with individual resistance $R$ if $\tau_b= \frac{\hbar^2
}{(2e)^2 RT}$ and $V_0 \rightarrow \infty$.
In order to average the solution of (\ref{langevin}) over the thermal noise,
$\zeta$, we use a generating functional.
For example, the average supercurrent in the array is given by
\beq
\langle I_{ij}\rangle = \int I_{ij} \exp
{\cal A}[S,\hat{S}]  {\cal D}S {\cal D}\hat{S}.
\label{I}
\eeq
Here the current $I_{ij} = \left( \frac{2e}{\hbar c}\right) \mbox{Im}
S_i^*J_{ij}S_j $ and
the action is
\beq
{\cal A} = \int dt \left[ \hat{S} \left(\tau_b\dot{S} + \frac{1}{T}
\frac
{\partial ({\cal H} + V)}{\partial S} \right) +
 \tau_b\hat{S}^2 \right ]
\label{A}
\eeq
where we have not included the terms that arise
from the Jacobian since they do not affect the long-time
response.\cite{dominicis78,sompolinsky82,jetp87}

We perform our calculations by resumming the terms in the $\frac{\cal H}{T}$
expansion of (\ref{I}) which are leading order in $\frac{1}{N}$;
this is a dynamical analogue of the high-temperature series expansion
previously used to study the static behavior of this array.\cite{us}
The crucial ingredients of this technique are the response ($G_{mn}(t,t') =
\langle s_m(t) \hat{s}_n(t')\rangle$)
and the correlation ($C_{mn}(t,t') = \langle s_m(t)s^*_n(t')\rangle$)
functions.
For $T > T_G$, these depend {\sl only} on the time-differences and thus can be
related by the Fluctuation-Dissipation Theorem
\beq
G_{ij}(t-t') = -  \frac{\partial D_{ij}(t-t')}{\partial t}\theta(t -
t')
\label{fdt}
\eeq
The leading diagrams (in $\frac{1}{N}$) for $G_{ij}(t-t')$ are displayed in
Fig. 2a.
The presence of the  ``constraining potential'' $V$ in the action (\ref{A})
results in finite higher-order irreducible single-site spin correlations, which
play the role of interaction vertices in this diagrammatic technique.
However the corrections to the response function shown in Fig. 2b are small in
$\frac{1}{N}$ in comparison with those in Fig. 2a.

We emphasize that, as in the static case,
the single-site response
is renormalized;
here we consider the local Green's function ($\tilde{G}(t-t')$)
that
is irreducible with respect to the $J_{ij}$ lines.
Possible self-energy corrections to $\tilde{G}(t-t')$ are shown in Fig. 2c
and will be discussed below.
Summing the geometric series shown in Fig. 2a, we obtain
\beq
\hat{G}_\omega =
\frac{1} {\tilde{G}^{-1}_\omega - \beta^2 (J^\dagger J)\tilde{G}_\omega}
\label{G}
\eeq
for the response function connecting wires of the same type
(horizontal/vertical).
The matrix $(J^\dagger J)_{ij}$ depends  only on the ``distance'' $i-j$
and acquires a simple form in Fourier space $(J^\dagger J)_p =
(J_0^2/\alpha) \theta(\alpha \pi - |p|)$; in this representation the
Green function becomes
\beq
G_\omega(p) = \frac{\theta(\alpha \pi - |p|)} {\tilde{G}^{-1}_\omega -
	\frac{(\beta J_0)^2}{\alpha} \tilde{G}_\omega}  +
	\frac{\theta(|p| - \alpha \pi)} {\tilde{G}^{-1}_\omega}.
\label{G_p}
\eeq
The static limit ($\omega = 0$)  of $T \tilde{G}^{-1}_\omega$  coincides with
the locator, $A(T)$, discussed previously \cite{us}; in the absence of
Onsager feedback terms $\tilde{G}^{-1}_0 = 1$.
Therefore we see in (\ref{G_p}) that there would be a static instability
at $\tilde{G}^{-1}_0 = G_c = \frac{\beta J_0}{\sqrt{\alpha}}$.
For $\Theta=(T-T_0)/T_0 \gg \sqrt{\alpha}$ all feedback effects are negligible.
Here the time-dependence of $\tilde{G}^{-1}_\omega$  is set by a microscopic
time-scale $\tau_b$ and $\tilde{G}^{-1}_b(\omega) = \beta A(T) - i \omega
\tau_b$; inserting this $\tilde{G}^{-1}_b(\omega)$ into (\ref{G_p}) we see
that the long-time behavior of $G_\omega(p)$ is dominated by the first term
which results in a long relaxation time $\tau=(2/a)\tau_b$ where $a = \beta
(A - \frac{J_0^2}{\alpha A}) \approx 2 \Theta $.
In this regime the single-site response function is
\beq
G(t) = \frac{\alpha}{2\tau_b} e^{-t/\tau} \hspace{0.5in} t \gg \tau_b.
\label{G(t)}
\eeq

At lower temperatures, $\Theta \lesssim \sqrt{\alpha}$, the feedback effects
become important; they modify $\tilde{G}^{-1}_0$  so that it approaches $G_c$
only asymptotically at $T \rightarrow 0$ and instead a first-order transition
occurs \cite{us}.
The retardation of the Onsager terms is also crucial and significantly
affects the
long-time behavior.
Qualitatively the resulting dynamical instability, described
below, is due to the time-dependence of the cavity field which
itself is determined by single-site susceptibilities;
the time-scale associated with the relaxation of
$G(t)$ increases continuously
due to feedback through the Onsager terms.
Formally the latter
introduce an additional frequency-dependent part of the local
response
\beq
\tilde{G}^{-1}_\omega=\tilde{G}^{-1}_b(\omega) - \Sigma_\omega
\label{tildeG}
\eeq
as a self-energy $\Sigma_\omega$ such that $\Sigma_0 = 0$ since we have
chosen our normalization so that $\tilde{G}^{-1}_0 = \beta A(T)$.

The on-site self-energy terms (see Fig. 2c) are the simplest
for $\alpha \ll 1$, and thus we will consider this regime.
We focus on the long-time response of this sytem which is
dominated by the first term in
(\ref{G_p}) when $ \tilde{G}^{-1}_0  \approx G_c $; its weight is proportional
to $\alpha$ (cf. (\ref{G(t)})).
Thus in the limit of $\alpha \ll 1$, the slowly decaying parts of $D(t)$ and
$G(t)$ scale with  $\alpha$ and the dominant self-energy contribution contains
the minimal number of these functions ($\Sigma^{(3)}$ in Fig. 2c)
and is given by
\beq
\Sigma_{\omega} = \Gamma^2 \int D^2(t) G(t) \left (e^{i\omega t} - 1 \right) dt
\label{Sigma}
\eeq
where $D$ and $G$ are single-site correlation and response functions
respectively. \cite{remark}
Here $\Gamma$ is the four-spin vertex; we neglect its transient
time-dependence and approximate it by its static value $\Gamma = -1$
which is determined by the high-temperature single-site
nonlinear susceptibility $\chi_3 = -\frac{1}{T^3}$.
The set of equations,(\ref{fdt}),(\ref{G_p}), (\ref{tildeG})
and (\ref{Sigma}), are sufficient to determine the response and
the correlation functions of the array.

\begin{figure}
\vspace{.3in}
\centerline{\epsfxsize=5cm \epsfbox{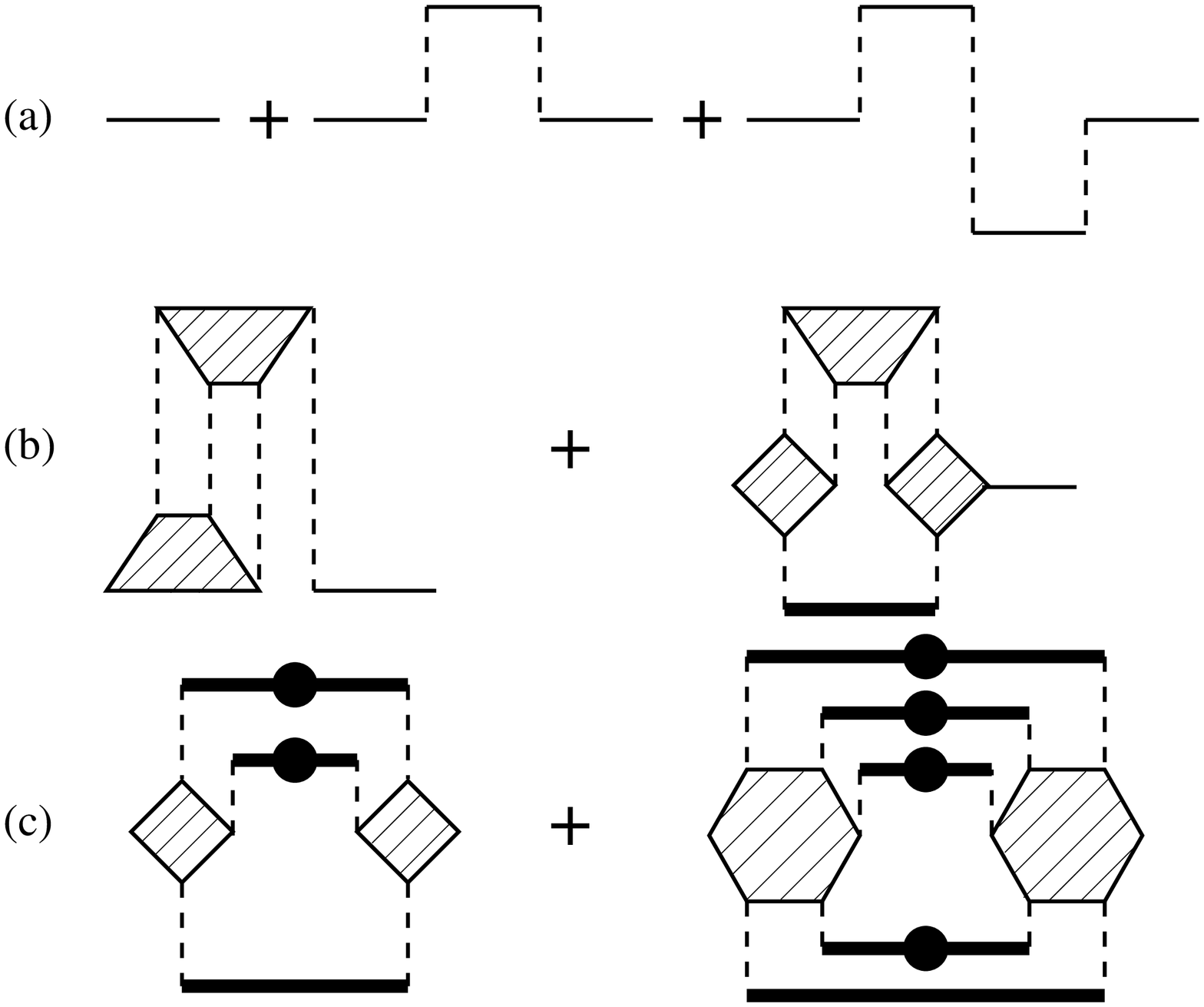}}
\vspace{.2in}
{Fig 2.
Diagramatic expansion for the response function $\hat{G}(t)$;
the dashed, solid and thick (with dot) lines are the coupling
matrix
$\cal J$, the
single site irreducible  ($\tilde{G}(t)$)
and the full response (correlation)  functions respectively.
a. Leading order in $1/N$.
b. Subleading order in $1/N$.
c. Leading terms in the expansion for $\tilde{G}(t)$; the first
diagram dominates at small $\alpha$.}
\end{figure}

Since we would like to detect a dynamical instability, we only
consider the long-time behavior of the response function, i.e.
\beq
G(t) = \alpha \int \frac{e^{-i\omega t}}{a - 2 (\Sigma_\omega +
i\omega \tau_b)} \left( \frac{d \omega}{2 \pi} \right).
\label{G(t)b}
\eeq
We make the ansatz
\beq
\Sigma_\omega = \frac{i\omega\tau_0/2}{1 - i\omega \tau_1}
\label{ansatz}
\eeq
and show that it solves the system of equations
(\ref{fdt},\ref{G_p},\ref{tildeG},\ref{Sigma}); in the process we also obtain
the time-scales $\tau_0$ and $\tau_1$ controlling the physical response.
We have chosen this form
of $\Sigma_\omega$ so that $\Sigma_0 = 0$ and $\Sigma(t) \sim
exp(-t/\tau_1)$.
Inserting this ansatz into (\ref{G(t)b}), we obtain
response and correlation functions that are simple exponentials
for long times; therefore the resulting self-energy (cf. (\ref{Sigma}))
is also exponential and thus of the same form as the initial ansatz.
Equating the parameters of (\ref{ansatz}) and (\ref{Sigma}),
we find that the condition for self-consistency at $a \rightarrow a_c$ is
\beq
\tau_1 = \frac{1}{3}\frac{\tau_b}{a - a_c},\;\;\; \tau_0 \cong 2 a_c \tau_1,
\;\;\; a_c = \frac{1}{3}(2\alpha)^{\frac{3}{4}}
\label{a_c}
\eeq
The resulting long-time part of the single-site response and the correlation
functions are
\beq
G(t) \!=\! \frac{2\alpha}{3a\tau_R}
	e^{-\frac{t}{\tau_R}}, \;
D(t) \!=\! \frac{2\alpha}{3a}
	e^{-\frac{t}{\tau_R}}
\label{G(t)c}
\eeq
where $\tau_R = 3\tau_1$, and we see from (\ref{a_c},\ref{G(t)c}) that the
longest
physical time-scale, $ \tau_R$, diverges continuously at
$\Theta_G = -\frac{3}{2} (2\alpha)^{1/4}$ (cf. Fig. 1). At $\Theta = \Theta_G$
the long-time part of $D(t)$ shown in (\ref{G(t)c})
becomes constant $q = (2\alpha)^{1/4}$ showing
that the Edwards-Anderson order parameter jumps  at the $\Theta_G$.
The resulting phase diagram is displayed in Figure 1; we note
that $T_G = (1 + \Theta_G)T_0$ occurs at a lower temperature
than $T_m$, where the last low-temperature
metastable states disappear\cite{us} as discussed above.

The response function $G(t)$ is a susceptibility with respect
to the field conjugate to $S = \exp{i\phi}$, and thus cannot
be measured directly. However, using $G(t)$ found above,
we can  determine the ac-response to a time-varying
physical magnetic field $H(t)$ which is experimentally
accessible.  We focus
on the total magnetic moment of the array generated by the
Josephson currents:
\beq
{\cal M} = \frac{1}{2} \left(\frac{2e}{\hbar c}
\right) l^2
\sum_{mn}
\langle S_m
{\tilde{\cal J}}_{mn} S_n \rangle
\label{M}
\eeq
where ${\tilde{\cal J}}_{mn}= imn{\cal J}_{mn}$ if $m$ and $n$
are indices referring to horizontal and vertical wires respectively.
We would like to determine the response in this magnetization
to a time-varying field; we use the fact that
${\cal M} = 0$ for static $H$ to
write
\beq
\frac{\partial {\cal M}(t)}{\partial H (t')} =
\left( \frac{2e}{\hbar c}\right)^2 l^2 {\mbox Re} {\mbox Tr}
{\tilde{\cal J}} {\hat G(t,t')} {\tilde{\cal J}} {\hat D(t',t)}.
\label{dMdH}
\eeq
In order to evaluate (\ref{dMdH}) we will need the response
function connecting wires of different type (horizontal/vertical)
\beq
\hat{G} = \hat {J} \frac{1}{ {\tilde{G}^{-2}} - \beta^2\hat
{J}^\dagger\hat {J}}
\label{hatG}
\eeq
and of the same type (cf. (\ref{G})).
We use the Fourier representation of $\hat {J}^\dagger\hat {J}$
and that of $\hat{G}$ and $\hat{D}$ to determine the ac-response,
$\frac{\partial {\cal M}(t)}{\partial H (t')}$,
in terms of the single-site response functions ( $L = Nl$):
\begin{eqnarray}
{\cal M}(t)= \! \left( \frac{2e}{\hbar c} \right)^2 \! \left(\frac{L^2}
{12}\right)^2 N \frac{J_0^2}{T}\frac{1}{\alpha^2} \left(
1-\frac{J_0^2}{A^2\alpha} \right)
\nonumber \\
\int_{-\infty}^t G_{t - t'} D_{t - t'} [H(t) \!-\! H(t')] \! dt'.
\label{mfinal}
\end{eqnarray}
We can insert the response and correlation functions found above
to determine the ac susceptibility
$\chi_\omega = \frac{\partial {\cal M}_\omega}{\partial H_\omega}$
which leads to
\beq
\chi_\omega = - \left( \frac{2e}{\hbar c}\right)^2 \frac{2N}{9}
\left(\frac{L^2}{12}\right)^2 \frac{J_0\sqrt{\alpha}}{a}
\frac{\omega}{\omega + 2i/\tau_R}
\eeq
where  $\tau_R$ is the longest time-scale of the response:
\begin{eqnarray}
\tau_R &\approx& \left\{
\begin{array}{ll}
	\frac{2\tau_b}{a(\Theta)} \hspace{0.5in}
	& \Theta > 0 \\
	\frac{\tau_b}{a_c}\frac{|\Theta_G|}{(\Theta-\Theta_G)}
	& \Theta-\Theta_G \ll |\Theta_G|
\end{array}
\right.
\label{tau_R} \\
a(\Theta) &=& \Theta + \sqrt{\Theta^2 + 2\alpha}
\label{a}
\end{eqnarray}
 From Eq. (\ref{tau_R}) we see that the divergent relaxation times are
directly observable in the physical a.c. magnetic response of the
array.
The zero frequency limit of the a.c. susceptibility jumps to a  finite
value at $T = T_G$, indicating the development
of a finite superconducting stiffness at the transition.
Therefore, the measurement of this a.c. response in a
fabricated array serves as a probe of glass formation.

In summary, we have presented a periodic model which displays
a dynamical transition where the system ``freezes'' into
one of an extensive number of metastable states.
Due to their large degeneracy, these states will not
be selected by a Boltzman weight under equilibrium
conditions.
This glass transition at $T_G$
is characterized by  diverging relaxation times and by
an accompanying jump in the Edwards-Anderson order parameter;
the phase diagram of the array is displayed in Figure 1.
It would be interesting to study the physical properties
of this array below the glass transition temperature
in its non-ergodic regime; in particular we expect
``memory'' effects in the form of an anomalous response
function and ``fingerprints'' of the individual
metastable states in its physical behavior.
Since any uncertainty in the position of
the wires
introduces randomness in this array, it also
offers the opportunity to study the crossover
between glasses with spontaneously-generated and
quenched disorder.

We thank D.M.Kagan for several useful discussions and for
a careful reading of the text.
M.V.F. acknowledges partial support
from the International Science Foundation, the Russian Government
(joint grant M6M300) and the Russian Foundation for Fundamental
Research (grant \#95-02-05720); M.V.F. and L.B.I. thank
NEC Research Institute for hospitality.

\end{multicols}

\begin{references}
\bibitem{remark} Strictly speaking the high-temperature expansions for $D(t)$
and $G(t)$ in (\ref{Sigma}) differ from those of the  correlation and response
function respectively; the former contain additional factors of
$(\beta^2 {\cal J}^\dagger {\cal J})$ that do not affect the long-time
behavior near $T_0$.
\end{references}
\end{document}